\documentclass{egpubl}

\usepackage{hpg21}
\usepackage[T1]{fontenc}
\usepackage{dfadobe}  

\usepackage{cite}

\ifpdf \usepackage[pdftex]{graphicx} \pdfcompresslevel=9
\else \usepackage[dvips]{graphicx} \fi

\usepackage{listings}
\usepackage{array}
\usepackage{enumitem}
\usepackage{algorithm} 
\usepackage{algpseudocode}

\usepackage[utf8]{inputenc}

\usepackage{color}

\input{listings-glsl.prf}

\definecolor{dkgreen}{rgb}{0,0.6,0}
\definecolor{gray}{rgb}{0.5,0.5,0.5}
\definecolor{mauve}{rgb}{0.58,0,0.82}

\title[Screen Space Indirect Lighting with Visibility Bitmask]
      {Screen Space Indirect Lighting with Visibility Bitmask}

\author[O. Therrien, Y. Levesque, G. Gilet]
{\parbox{\textwidth}{\centering Olivier Therrien$^{1}$\thanks{therrien.olivier@cdrin.com}
         Yannick Levesque$^{2}$\thanks{levesqueyannick@cgmatane.qc.ca} Guillaume Gilet$^{3}$\thanks{guillaume.gilet@usherbrooke.ca} 
        }
        \\
{\parbox{\textwidth}{\centering $^1$CDRIN, QC, Canada\\
        $^2$Cégep de Matane, QC, Canada\\
        $^3$ University of Sherbrooke\\
   }
}
}
\begin{document}

\teaser{
 \includegraphics[width=\linewidth]{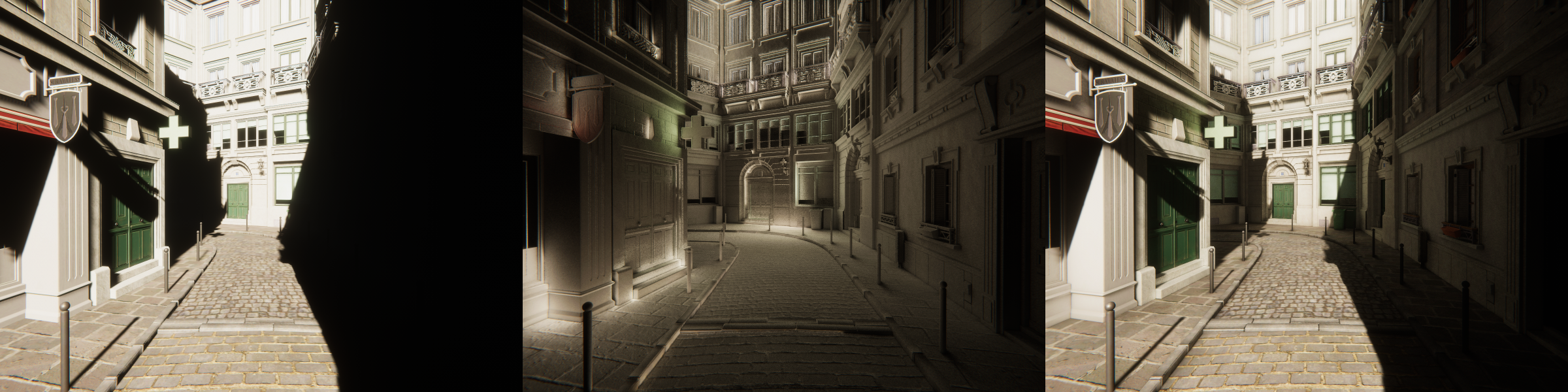}
 \centering
  \caption{Left: Direct illumination of the scene. Middle: Indirect lighting produced by our method (without texture). Right: Final frame rendered with our method, exhibiting directionally occluded ambient lighting, and a GI bounce that avoids typical thin surface artifacts.}
\label{fig:teaser}
}

\maketitle
\begin{abstract}
   Horizon-based indirect illumination efficiently estimates a diffuse light bounce in screen space by analytically integrating the horizon angle difference between samples along a given direction. Like other horizon-based methods, this technique cannot properly simulate light passing behind thin surfaces. We propose the concept of a visibility bitmask that replaces the two horizon angles by a bit field representing the binary state (occluded / un-occluded) of N sectors uniformly distributed around the hemisphere slice. It allows light to pass behind surfaces of constant thickness while keeping the efficiency of horizon-based methods.  It can also do more accurate ambient lighting than bent normal by sampling more than one visibility cone. This technique improves the visual quality of ambient occlusion, indirect diffuse, and ambient light compared to previous screen space methods while minimizing noise and keeping a low performance overhead.
   \\


\end{abstract}  
\section{Introduction}

\begin{figure}[htb]
  \centering
  \includegraphics[width=.66\linewidth]{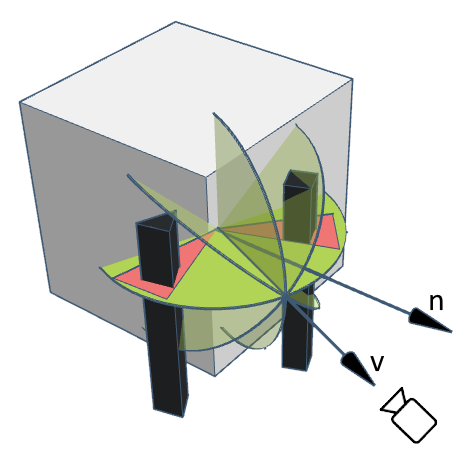}
  %
  %
  \caption{\label{fig:Slice3D}
           3D view of the scene, centered on the shaded pixel. GTAO/HBIL generates a set of hemisphere slices (in green) in various directions around the view vector \textit{v}. Occluders (in black) intersect some of the slices, producing occlusion cones (in red).}
\end{figure}

Indirect diffuse lighting is challenging to compute in real-time. Screen space approximations can be attractive as they reduce the dimensionality of the problem and make the execution cost constant regardless of the scene’s geometric complexity. Modern Screen Space Global Illumination (SSGI) implementations often gather indirect light by doing ray marching on screen pixels similar to Screen Space Reflections (SSR) \cite{sousa2011secrets}. This approach tends to generate a lot of noise because it implies the numerical integration of irradiance over the entire hemisphere around the surface. Horizon-Based Indirect Illumination  (HBIL) \cite{mayaux2018}, based on Horizon-Based Ambient Occlusion (HBAO)  \cite{bavoil2008image, bavoil2011horizon} and Ground Truth Ambient Occlusion (GTAO) \cite{jimenez2016practical}, improve the efficiency by numerically integrating over a set of directions around the view vector \textit{v} (Figure \ref{fig:Slice3D}) while doing analytic integration of the horizon angle difference between samples.

\begin{figure}[htb]
  \centering
  \includegraphics[width=1.0\linewidth]{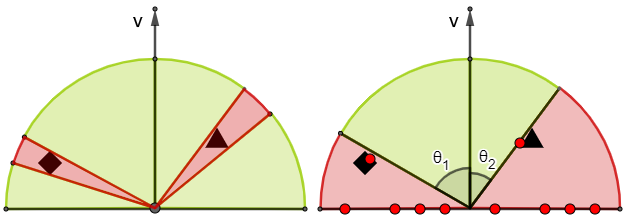}
  %
  %
  \caption{\label{fig:SliceGTAO}Side view of one slice centered on the view vector \textit{v}. Left: Ground truth scene with multiple occluders (in black) producing multiple visibility cones (in green). Right: GTAO takes a fixed number of samples (red dots) in the depth buffer on both sides of the hemisphere to find highest elevation angles $\theta_1$ and $\theta_2$.}
  
\end{figure}

\begin{figure}[htb]
  \centering
  \includegraphics[width=1.0\linewidth]{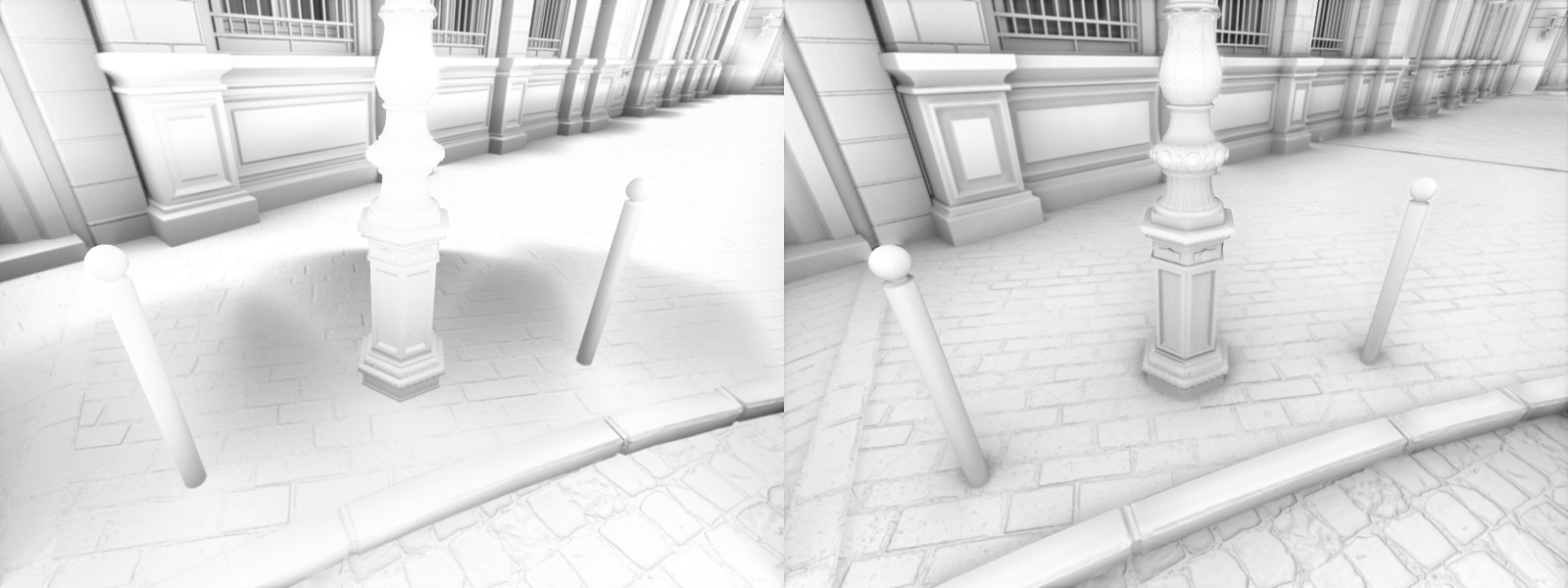}
  %
  %
  \caption{\label{fig:AOHalos}
           Left: GTAO produces halos around the poles. Right: Our method is able let light pass behind the poles without introducing halos artifacts.}
\end{figure}

Fundamentaly, the core principle of these methods lies in the estimation of the scene local geometry around each shading sample by relying on readily-available screen-space information, such as the discrete depth buffer. However, such information is by essence discrete and incomplete, and must be reconstructed. 
All those techniques evaluate Ambient Occlusion, the modulation of indirect irradiance due to local geometry, in screen space from a single layer depth buffer, and assume infinite surface thickness by treating it as a height-field (see Figure \ref{fig:SliceGTAO}). While this is a valid assumption in some cases, not knowing what the real geometry looks like, it causes halos and over-darkening around thin surfaces (see figure \ref{fig:AOHalos}). Falloff heuristics are used to mitigate those artifacts but fail when using a large sampling radius.

Our proposed method rejects the assumption that the depth buffer is strictly a height-field and models the behavior of light passing behind surfaces. Additionally, to preserve performance, we want to avoid explicitly tracing new rays from scratch to adequately sample multiple elevation angles. To do so, we introduce the concept of visibility bitmask, which is essentially a discretization of the hemisphere slice in \textit{$N_b$} sectors, that allows us to approximate the tracing of \textit{$N_b$} rays at the same performance cost as one horizon search. 

A common limitation of single layer depth buffers is that occluded surfaces are not represented, which can cause missing occlusion or lighting. Approaches like Deep G-Buffer \cite{mara2016deep} or Multi-layer SSRT \cite{hofmann2017hierarchical} alleviate this issue by storing multiple layers or sample per pixel, to provide more information on background surfaces. As rendering multiple layers is very expensive, we chose to limit ourselves to only one layer. This means that some artifacts caused by inaccurate background surface estimation will remain, but using a constant thickness \textit{t} at each depth sample with a visibility bitmask greatly improves quality around thin surfaces compared to horizon-based techniques. In this study, we show that visibility bitmasks can tremendously reduce noise in the image compared to SSR-like tracing, while handling the light passing behind surfaces much more accurately than horizon-based techniques. We also show that using a visibility bitmask to sample ambient light gives a more precise ambient estimation than traditional methods of sampling along the surface normal or even bent normal.


The main contributions of this paper revolve around the introduction of a visibility bitmask in traditional SSGI methods, retaining the efficiency and noise reduction qualities of horizon sampling, while handling light passing behind surfaces of constant thickness. We demonstrate the capabilities of our method through several SSGI applications, such as ambient occlusion, directional ambient occlusion and indirect diffuse lighting.


\section{Previous Work}
Estimating ambient occlusion and indirect diffuse lighting from screen space information is a well-known idea. It mainly stems from the seminal Screen Space Ambient Occlusion method \cite{mittring2007finding}, approximating ambient occlusion by sampling random points in the depth buffer in a circle around each pixel and has been thoroughly extended over the years \cite{mittring2007finding, mcguire2011alchemy, mcguire2012scalable}. The keypoint of these methods is to estimate the local geometry around a sample using the incomplete information contained in the various buffers (geometrical normals, depth...) while maintaining high rendering performance. 

Reconstruction of local geometry can be improved by gathering more information from the scene during the rendering passes. Reflective Shadow Maps (RSM) \cite{dachsbacher2005reflective} approximate indirect diffuse lighting coming from a point light source using essentially a G-Buffer generated from the light's view point. This technique is costly so in practice it's usage is limited to less than a handful of light simultaneously, and doesn't take into account indirect light occlusion. Deep Screen Space \cite{nalbach2014deep} adaptively tessellate scene geometry into an unstructured surfel cloud used for rendering different effects like AO, GI and more. It bypasses major screen-space limitations like hidden surfaces and under-sampling of oblique geometry, but is expensive to compute and cannot handle indirect light occlusion between surfels. \\
More recently, Stochastic-Depth Ambient Occlusion (SDAO) \cite{vermeer2021stochastic} introduced the notion of stochastic depth map, capturing multiple scene layers per pixel at random. This technique is effective at detecting hidden surfaces, but, since it's used in conjunction with HBAO \cite{bavoil2008image}, it doesn't prevent over-darkening around thin objects. However, while these methods improve reconstruction of local geometry, they are more computationally expensive than single-layer approaches, both during sample capture (by forgoing early-z optimization) and reconstruction (by having to evaluate multiple layers). \\

Improving the quality of the reconstruction from a single layer is a difficult problem that has been widely studied. Alchemy ambient obscurance \cite{mcguire2011alchemy} is based on a similar approach than SSAO and improves robustness and artistic control, with the follow up Scalable Ambient Obscurance (SAO) \cite{mcguire2012scalable} that also improves performance. Horizon-Based techniques \cite{bavoil2008image, bavoil2011horizon} generates high quality results with low amount of noise by sampling elevation along a set of directions but causes over-darkening around thin surface. Several methods focus on improving performance, such as Line Sweep Ambient Obscurance (LSAO) \cite{timonen2013line}, which pre-caches sample information along azimuthal lines in GPU shared memory and reuses the same samples to shade multiple pixels. More recently, Ground Truth Ambient Occlusion\cite{jimenez2016practical} improved the accuracy of HBAO by making it match a path-traced reference, and support a multi-bounce occlusion approximation.\\

These techniques has been derived to propose more advanced indirect illumination features, taking advantage of the local geometry reconstruction. Silvennoinen \textit{et al.}\cite{silvennoinen2015multi} added support for indirect lighting in real-time via an SSGI implementation  using LSAO as a basis. This method is approximative because only one color sample is taken per horizon highpoint, and it doesn't take into account partial occlusion that could have occurred along the way. Other SSGI variants like Screen Space Ray Tracing Global Illumination (SSRTGI) \cite{shergin2017superposition} use an SSR-like technique to sample GI at the ray hit location. However, this approach introduces a lot of noise which is difficult to remove without over-blurring. Another technique known as HBIL (which is based on HBAO and GTAO) showed how to compute GI accurately from multiple samples by weighting the sample contribution by the angle difference relative to the previous sample. While this method gives accurate results within the visibility cone, it's based on the assumption that the depth buffer is a height field, and it cannot take into account light bounces that would pass behind surfaces. \\

Finally, Bitmask Soft Shadows (BMSS) \cite{schwarz2007bitmask} determine the visibility of an area light source with a bit field where each bit tracks the visibility of a sample point on the light source. Our method solves a slightly different problem but nonetheless shares many similarities regarding surface shape estimation from a depth map and addresses overlapping sample visibility in the same way using a bitmask.



\section{Proposed Algorithm}
In this section, we present how our method propose to improve the quality of reconstruction of local geometry, especially in the case of thin surfaces, by treating the depth buffer as a set of unconnected samples each associated with an arbitrary thickness. 
\subsection{Ambient Occlusion}

Ambient Occlusion (AO) \cite{zhukov1998ambient} is a non-physically based lighting approximation of global illumination that assumes that the scene is lit by uniform ambient lighting and that all objects are occluders. It’s a very common effect in real-time applications because it can be computed efficiently in screen space and adds a lot of perceived realism to the scenes. It can be expressed as a an estimation of the visibility function $V$ in the hemisphere around each sample :
\begin{equation}
    AO = 1 - \frac{1}{\pi}\int_{\Omega} V(p,\omega) (n_p . \omega) d\omega 
    \end{equation}
By using a parameterization of the hemisphere, it can be decomposed into :
     \begin{equation}\label{outerintegral}
     AO = 1 - \frac{1}{\pi}\int_{0}^{\pi} AO_2(\phi)d\phi\\
     \end{equation}
     and 
      \begin{equation}\label{innerintegral}
    AO_2(\phi) = \int_{0}^{\pi}V(p,\theta,\phi) \cos\theta \sin \theta d\theta
\end{equation}

In practice, the integral of equation \ref{outerintegral} is computed using Monte Carlo integration over a few slices. Method such as GTAO and HBAO propose an analytic solution of equation \ref{innerintegral} by estimating, through depth sampling, two horizons $\theta_1$ and $\theta_2$.

In our proposed algorithm, the two horizon angles $\theta_{1}$ and $\theta_{2}$ of GTAO are replaced by a bit field of size \textit{$N_b$}, representing the binary state (occluded/un-occluded) of \textit{$N_b$} visibility sectors uniformly distributed around the hemisphere slice. Samples are still taken on each side of the view vector \textit{v}, but the bit field is centered on projected normal \textit{n} :



\begin{equation}\label{innerintegral2}
AO_{2}(\phi_i) \approx \frac{1}{N_b} \sum_{j=1}^{N_b} V(\phi_i,\theta_j)
\end{equation}


Each sample taken along the hemisphere slice will determine a potential occluder and impact the visibility function $V$ of the given sector. To determine the occluded state of a sector, we consider each sample as a local thin geometry having a thickness $t$, acting as an occluding geometry between angles $\theta_{f}$ and $\theta_{b}$. The activation of a sector depends on the \textit{hit criterion} which ensures sufficient overlap of these angles with the sector to get registered. $\theta_{f}$ is equivalent to \textit{$\theta$} used in GTAO (directly at the sample), and $\theta_{b}$ depends on sample thickness \textit{t}, (figure \ref{fig:SliceGTAO}). For the proposed algorithm, we used the round criterion which requires the sector to be half covered by the sample. The pseudo-code at line 15 to 17 in Algorithm \ref{alg:ALG1} shows how $(\theta_{f}, \theta_{b})$ are inferred from sample position and thickness. All occluded sectors are set at once, making the algorithm perform in O(1) for any sector count. Note that we must convert the angles from cosine space to angular space for the samples to be properly distributed around the hemisphere.

\begin{figure}[htb]
  \centering
  \includegraphics[width=0.6\linewidth]{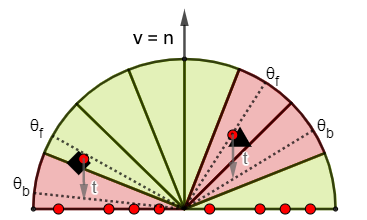}
  %
  %
  \setlength{\belowcaptionskip}{-100pt}
  \caption{\label{fig:SliceVS}
           The hemisphere is divided into $N_b$ uniform sectors that can be either occluded (in red) or un-occluded (in green). $\theta_{b}$ is derived from $\theta_{f}$ and thickness t. Sectors that are at least half covered by the $(\theta_{f}, \theta_{b})$ pair get occluded by the sample. We assume \textit{v} and \textit{n} are aligned in this diagram for simplicity.}
\end{figure}

This enables fast directional occlusion and partial integration, at the cost of precision. In the depth buffer, we take a fixed number of azimuthal directions around each pixel and sample along these directions to find ($\theta_{f}$,$\theta_{b}$) pairs that can be integrated into the hemisphere slice (see Figure \ref{fig:SliceVS}). Like GTAO we distribute the occlusion integral spatially and temporally to increase the number of effective samples.

\begin{figure}[tbp]
  \centering
  \includegraphics[width=1.0\linewidth]{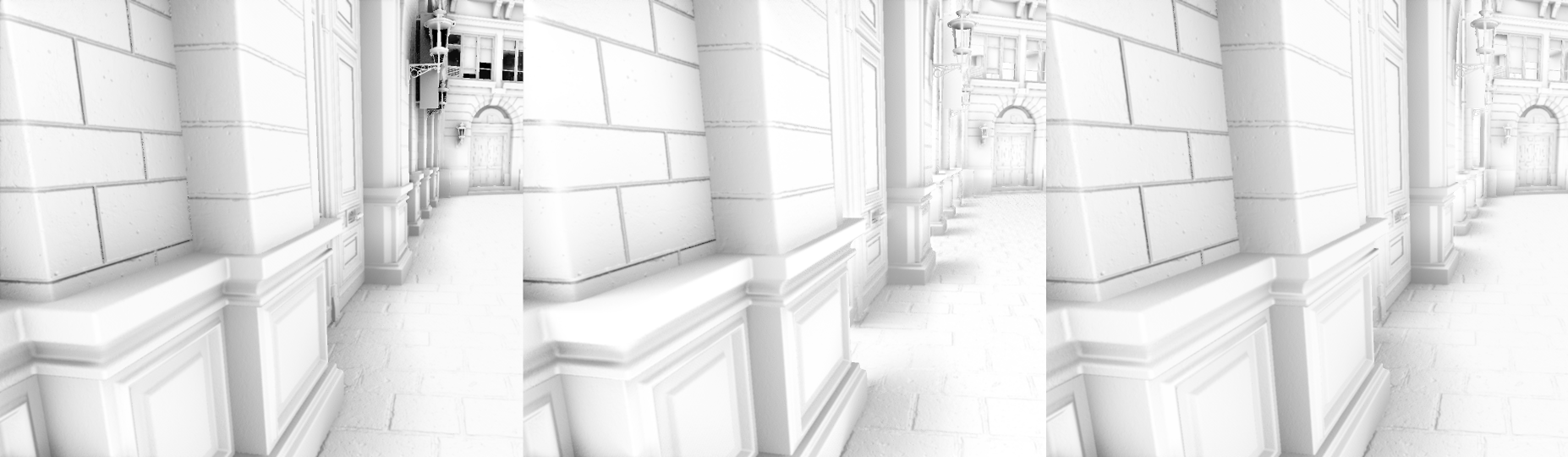}
  \caption{\label{fig:AOThicknessGaps}%
          Left: GTAO using horizon angles without falloff term exhibits no light leaks. Middle: our method using visibility bitmasks with fixed thickness causes light leaks at depth discontinuities. Right: GTAO using horizon angles with falloff term also causes some light leaks at depth discontinuities.}
\end{figure}

\begin{figure*}[tbp]
  \centering
  \includegraphics[width=1.0\linewidth]{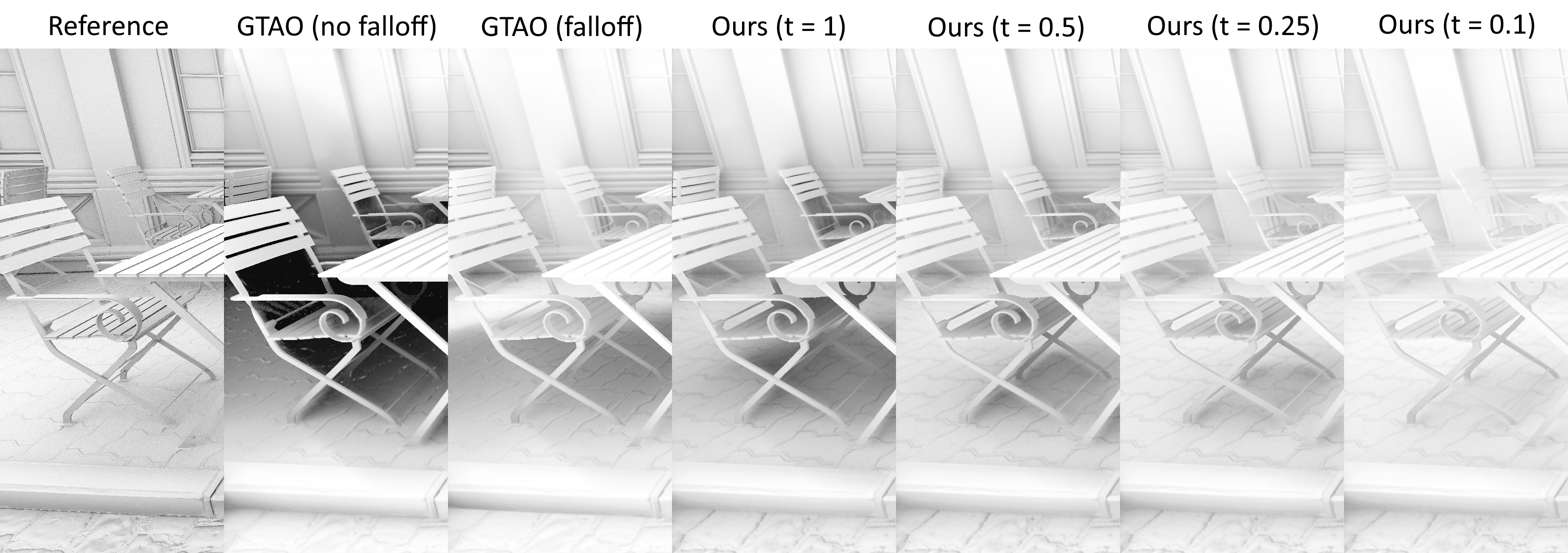}
  \caption{\label{fig:AOThickness}%
          From left to right: Ray tracing reference, GTAO without falloff, GTAO with falloff, visibiliy bitmask using a thickness of 1, 0.5, 0.25, 0.1. All methods use a radius of 2.}
\end{figure*}

The choice of \textit{t} has a big impact on visual fidelity. GTAO without falloff is equivalent to using an infinite value for \textit{t}. Ideally we would want to use the per-pixel surface thickness of objects as value for \textit{t}, unfortunately this is more expensive to compute in real-time and impossible to know precisely using a single layer depth buffer. We propose using a small constant value since artifacts of over-occlusion around thin objects are much more noticeable than light leaks behind thick objects. This is because bigger objects usually occlude completely what is behind them so a light leak can still look plausible, whereas over-occlusion around thin objects is directly visible (see Figures \ref{fig:AOThicknessGaps} and \ref{fig:AOThickness}). Using a fixed world-space thickness can cause an over-attenuation of occlusion for objects far away from the camera, so we give the option to increase \textit{t} linearly over the distance to counter this effect. This causes a slight change in occlusion when the camera moves, but is barely noticeable and fixes the problem effectively. Finding an efficient heuristic to estimate an accurate thickness for each sample would further improve the accuracy of the method but remains a difficult problem we leave for future work.
\begin{algorithm}
	\caption{Generate AO and GI using visibility bitmasks}\label{alg:ALG1} 
	\begin{algorithmic}[1]
	    \State $t\gets$ constant thickness
	    \State $N_{b}\gets$ bitmask size
	    \State $p\gets$ view space fragment position
	    \State $n_{p}\gets$ view space fragment normal
	    \State $r\gets$ projected radius onto image plane
        \State Determine stepsize as $r / (N_{s} + 1)$ 
        \State Determine directions with random offset
        \State $AO, GI \gets 0$
        \For{direction $d_{i}$ where $i = 0$ to $N_{d}$}
            \State $t_{p} \gets$ slice plane tangent vector in direction $d_{i}$
            \State $t_{\theta} \gets$ angle of $t_{p}$ with XY-plane
            \State Bitmask $b_{i} \gets$ 0
            \For{step $s_{j}$ where $j = 0$ to $N_{s}$}
                \State Front sample $s_{f} \gets$ view-space position at step j
                \State Back sample $s_{b} \gets s_{f} - \frac{p}{\|p\|} t$
                \State $\theta_{f}$, $\theta_{b} \gets$ angles of $s_{f}$ and $s_{b}$ on XY-plane
                \State $\theta_{min}$, $\theta_{max} \gets \min(\theta_{f}, \theta_{b}), \max(\theta_{f}, \theta_{b})$
                \State $a$, $b \gets \lfloor \frac{\theta_{min} + \frac{\pi}{2}}{\pi}N_{b} \rfloor$, $\lceil\frac{\theta_{max}-\theta_{min} + \frac{\pi}{2}}{\pi}N_{b}\rceil$
                \State $b_{j} \gets 2^{b}-1 \ll a$
                  
                \State $c_{j} \gets$ direct lighting at step $j$ (from GBuffer)
                \State $n_{j} \gets$ normal at step $j$ (from GBuffer)
                \State $l_{j} \gets \frac{s_{f} - p}{\|s_{f} - p\|}$
                \State $GI \gets GI + \frac{\Call{countbits}{b_{j} \mathbin{\&} \ensuremath{\mathord{\sim}}b_{i}}} {N_{b}} c_{j} (n_{p} \cdot l_{j}) (n_{j} \cdot -l_{j}) $
                
                
                \State $b_{i} \gets b_{i} \mathbin{|} b_{j}  $
              
            \EndFor
            \State $AO \gets AO + 1 - \Call{countbits}{b_{i}} / N_{b}$ 
            
        \EndFor
        \State \Return $AO / N_{d}$, $GI / N_{d}$
	\end{algorithmic} 
\end{algorithm}

The following sections explain the usage of the above-described core algorithm in implementing ambient occlusion, directionally occluded ambient lighting, and indirect diffuse bounce.

\subsection{Directionally Occluded Ambient Lighting}

%

\begin{figure}[htb]
  \centering
  \includegraphics[width=0.6\linewidth]{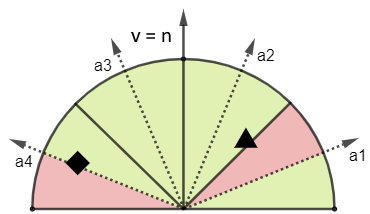}
  %
  %
  \caption{\label{fig:SliceAmbient}
           In this example, 4 ambient samples per slice per pixel are taken so the hemisphere is divided into 4 subregions. Vectors $a_{1}$, $a_{2}$, $a_{3}$, $a_{4}$ are generated at the center of their respective subregion.}
\end{figure}

Ambient lighting in real-time applications is usually sampled using the surface normal, but tends to give poor results because it doesn't take into account the directional occlusion of lighting. Screen space bent normal \cite{klehm2011bent} addresses this problem by sampling towards the largest non-occluded direction, but is limited to a single ambient direction per pixel and does not handle thickness properly. Visibility bitmasks can improve this by weighting the ambient lighting in a given direction by the directional occlusion while allowing light to pass behind surfaces. To this end, we divide the hemisphere into as many subregions as the number of ambient samples, generating a sampling direction vector at the center of each subregion. The ambient source is then sampled with this vector and multiplied the lighting intensity by the amount of un-occluded sectors over the total sector count (see Figure \ref{fig:SliceAmbient}). It made the ambient light color vary smoothly according to changes in occlusion directionality.

\subsection{Indirect Diffuse}
\begin{figure}[htb]
  \centering
  \includegraphics[width=1.0\linewidth]{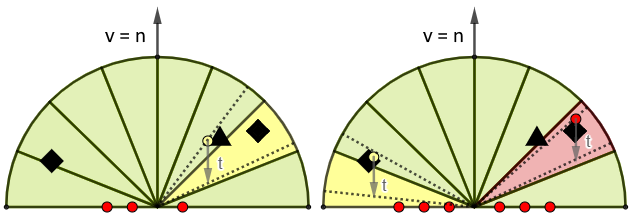}
  %
  %
  \caption{\label{fig:SliceGI}
           Left: The yellow sample intersects one un-occluded sector and can contribute lighting equivalent to one over the total number of visibility sectors. The sector is set to an occluded state for subsequent samples. Right: Sampling continues and a new object on the right is found, but it intersects an already occluded sector, so it cannot contribute lighting. The yellow sample on the left crosses an un-occluded sector and can contribute.}
\end{figure}

Indirect diffuse lighting is the bouncing of light on nearby surfaces. It’s traditionally expensive to compute accurately, even in screen space. HBIL can do it efficiently but fails to account for light passing behind surfaces. Figure \ref{fig:SliceGI} shows how we computed this effect with better thickness handling using visibility bitmasks. Samples were taken along the slice direction and detected (in O(1)) how many un-occluded sectors are covered to estimate lighting contribution. These sectors were then set to an occluded state (also in O(1)) to handle partial or total occlusion of light coming from subsequent samples. The more the visibility sectors, the more precise the estimation of lighting and occlusion. It was observed that 32 sectors gave good quality and makes the bit field fit nicely within a single unsigned integer. 

The pseudo-code at line 23 in Algorithm \ref{alg:ALG1} shows how the lighting contribution of a sample is implemented, using the number of occluded zones by the current sample. If one or more sectors are covered, the sample contributed lighting and the light buffer is sampled at the sample location. Then \textit{n} $\cdot$ \textit{l} and \textit{$n_{l}$} $\cdot$ \textit{l} are computed and used for weighting light intensity. The light is furthermore weighted by the occluded sector count over the total sector count.


\section{Results and evaluation}

\subsection{Ambient Occlusion}

\begin{figure*}[tbp]
  \centering
  \includegraphics[width=.8\linewidth]{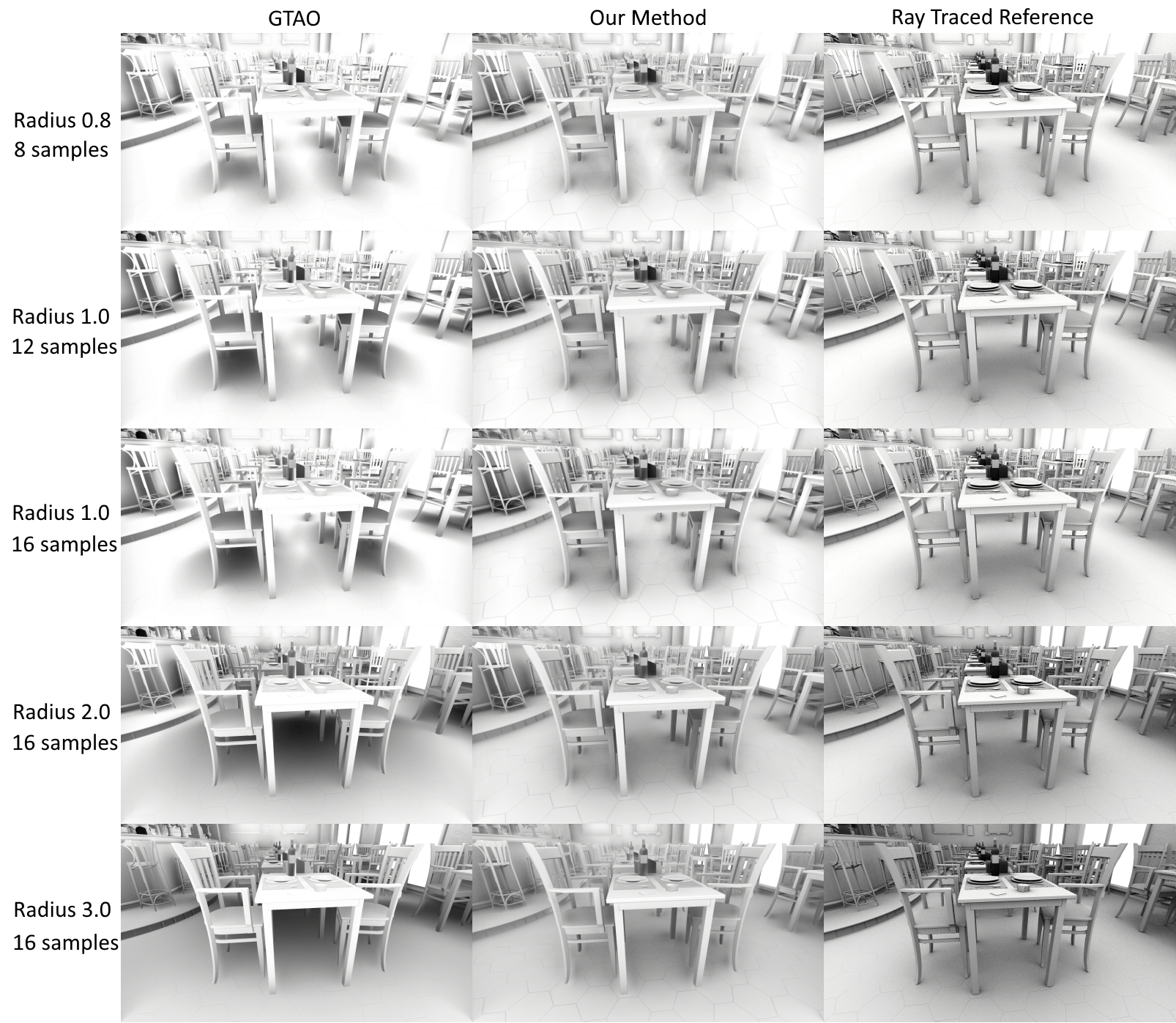}
  \caption{\label{fig:AOParam}%
           Comparison of GTAO, our method based on visibility bitmasks, and a ray-traced AO reference at different radius and sample count values.}
\end{figure*}

In this subsection, the core algorithm will be demonstrated conjointly with AO, as it’s much easier to discern the properties of visibility bitmasks in this mode than using indirect diffuse. The renders in Figure \ref{fig:AOParam} are produced by our extension of Unity’s GTAO implementation, where the two horizon angles $\theta_{1}$ and $\theta_{2}$ have been replaced by a single visibility bitmask.

It highlights how easy it is to implement our method on top of an existing horizon-based technique, and how this single modification can dramatically enhance visual quality compared to a ray-traced reference. The Lumberyard Bistro scene has been chosen because it contains a lot of thin and shallow surfaces that are typically a problem with horizon-based techniques, but are improved by visibility bitmasks. All benchmarks use one hemisphere slices per pixel jittered over multiple frames.

The radius parameter (Figure \ref{fig:AOParam}) is the radius of the hemisphere aligned to the screen in world units. Wider hemispheres will cover wider regions of the screen, casting farther-reaching occlusion. A wide radius is problematic for GTAO because the single cone approximation tends to cast too much occlusion in regions enclosed by thin objects. Even around not-so-thin objects, a wide radius tends to produce a blurry occlusion blob that does not capture fine detail. In contrast, our method (with the exact same samples) is able to let light pass behind surfaces, avoiding over-occlusion and capturing a lot of small details. 

\begin{figure}[htb]
  \centering
  \includegraphics[width=1.0\linewidth]{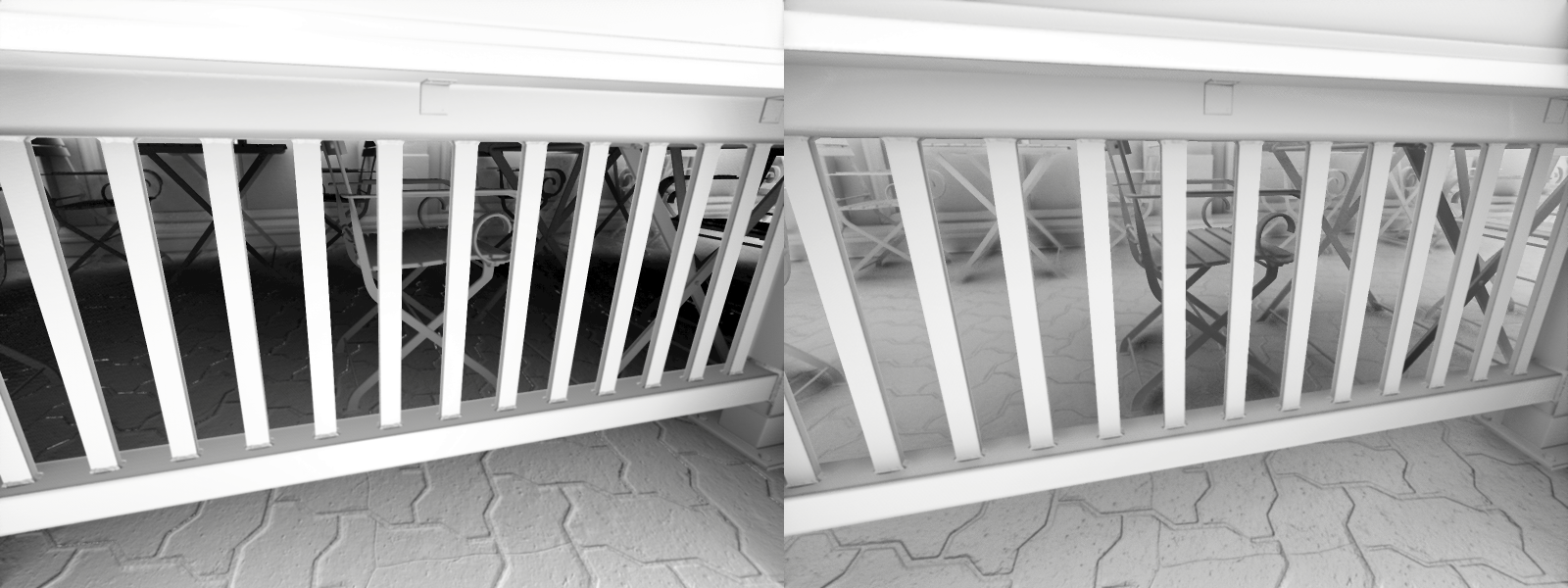}
  %
  %
  \caption{\label{fig:AOFence}
           Left: GTAO exhibit too much occlusion behind the bars. Right: Our method is able let light pass behind the bars, minimizing the over-occlusion artifact.}
\end{figure}

Figure \ref{fig:AOFence} demonstrates an even more difficult case for GTAO where most of the objects are behind a fence. The light gets trapped behind the bars instead of passing by, causing a lot of over-occlusion. Our method is able to handle this situation much better. When using a wide radius, GTAO has a tendency to produce halos around objects (Figure \ref{fig:AOHalos}). Our method doesn't have this problem, and capture more geometric and normal details.

The number of samples (Figure \ref{fig:AOParam}) indicates the number of fetches taken in the depth buffer along one horizon side. Therefore, for one slice, the actual number of samples taken is twice that number. A wider radius causes the samples to be sparser on-screen, so it's typical to increase sample count for a wider radius to maintain the same sampling density. A low sampling density increases the likelihood of missing potential occluders, especially if they are thin on-screen. Increasing the number of samples has a big impact on performance, so it’s a tradeoff.

Performance depends primarily on the number of samples taken and the radius of the effect. A wider radius increases the probability of occurrence of a cache miss (sample not being present in the cache) and consequently can lower performance. The execution time of the technique tends to scale linearly with the number of samples. Table \ref{table:1} compares the performance of the horizon-based GTAO implementation versus our method using visibility bitmasks. Both techniques are composed of a sampling pass and a denoising pass. Only the results of the sampling pass are included in the table since it’s the only one impacted by our method. The denoising pass has a constant cost of 0.3 ms in 1080p. It can be observed that our method has a modest impact on performance around 0.01-0.02 milliseconds, with a fixed ALU overhead of about 15 GPU instructions per sample. Increasing the radius masks this overhead as the execution becomes bandwidth-limited.

\begin{table}[h!]
\centering
\begin{tabular}{||c c c c||} 
 \hline
 Radius & Sample Count & GTAO & Our Method \\ [0.5ex] 
 \hline\hline
 0.8 & 8 & 0.49 ms & 0.51 ms \\ 
 \hline
 1 & 12 & 0.75 ms & 0.77 ms \\
 \hline
 1 & 16 & 0.95 ms & 0.97 ms \\
 \hline
 2 & 16 & 1.12 ms & 1.13 ms \\
 \hline
 3 & 16 & 1.12 ms & 1.13 ms \\
 \hline
\end{tabular}
\caption{Render time of the sampling pass for GTAO and our method with various radius and sample parameters, at 1920x1080, with 32 visibility sectors per hemisphere slice. Benchmarks are done on an Nvidia RTX 2080 GPU.}
\label{table:1}
\end{table}

Another parameter that impacts image quality is the number of visibility sectors. When the sector count is too low, banding artifacts can appear, particularly around thin objects. The proposed implementation used 32 visibility sectors because it just crossed the threshold where the artifacts became almost invisible. Additionally, it nicely fits into a single unsigned integer which gives good performance on the GPU. By contrast, a 128 bits version require four unsigned integers and the use of vector instructions, which limits the amount of instruction packing that the compiler could do, negatively impacting the performance. A performance overhead of around 5-10\% was observed with 128 visibility sectors compared to 32.



\subsection{Directionally Occluded Ambient Lighting}

\begin{figure}[htb]
  \centering
  \includegraphics[width=1.0\linewidth]{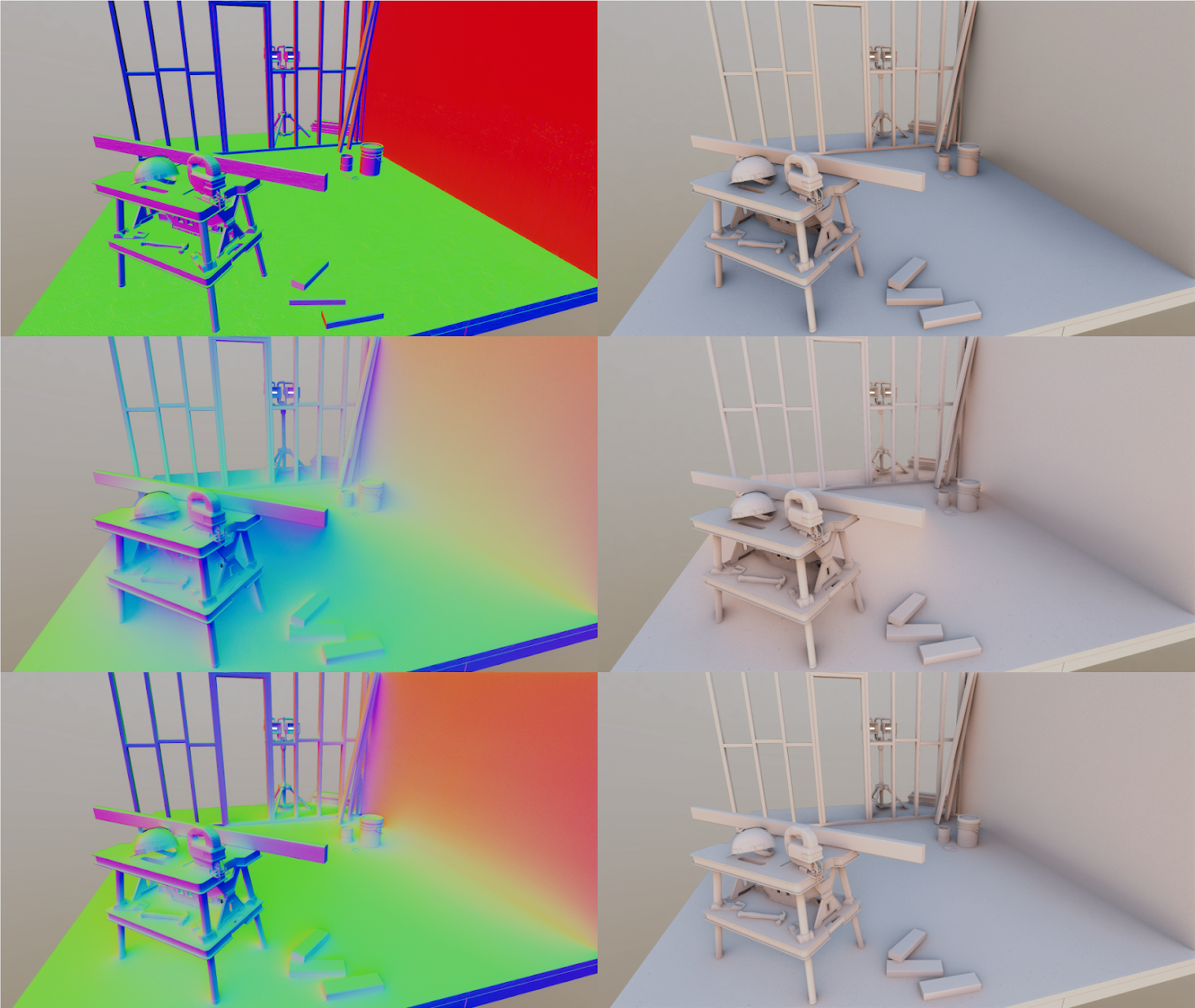}
  \caption{\label{fig:Ambient}%
          Comparison of ambient lighting using the surface normal, bent normals, and visibility bitmasks. The average sampling normal direction is shown on the left to make the difference more visible. Top row: G-Buffer normals. Middle row: Bent normals. Bottom row: visibility bitmask.}
\end{figure}

In this subsection, we compare different ambient sampling strategies and show how they can dramatically influence the resulting lighting and occlusion. Figure \ref{fig:Ambient} shows the average normal on the left, and the resulting ambient lighting on the right for each strategy. In most 3D applications, ambient lighting is sampled in the direction of the surface normal. This approach does not take into account the fact that some light could be occluded in some direction and tend to make ambient lighting change sharply with the scene geometry. A better approach is to use a screen space bent normal per pixel that is modulated according to nearby occlusion. It points towards the direction of incoming light and gives a smoother result. However, it cannot handle multiple light directions and will misrepresent the ambient lighting of surfaces enclosed by thin objects. 

Our approach used visibility bitmasks to take multiple samples along each hemisphere slice in the directions that were not occluded. Doing so allow ambient light to pass behind surfaces, giving smooth lighting from multiple directions. It's also worth noting that each ambient sample is weighted according to the occlusion in that specific direction. If the ambient color had been simply averaged out and then multiplied by ambient occlusion, a lot of the color variation in the lighting would have been lost.

\begin{figure*}[htb]
  \centering
  \includegraphics[width=.88\linewidth]{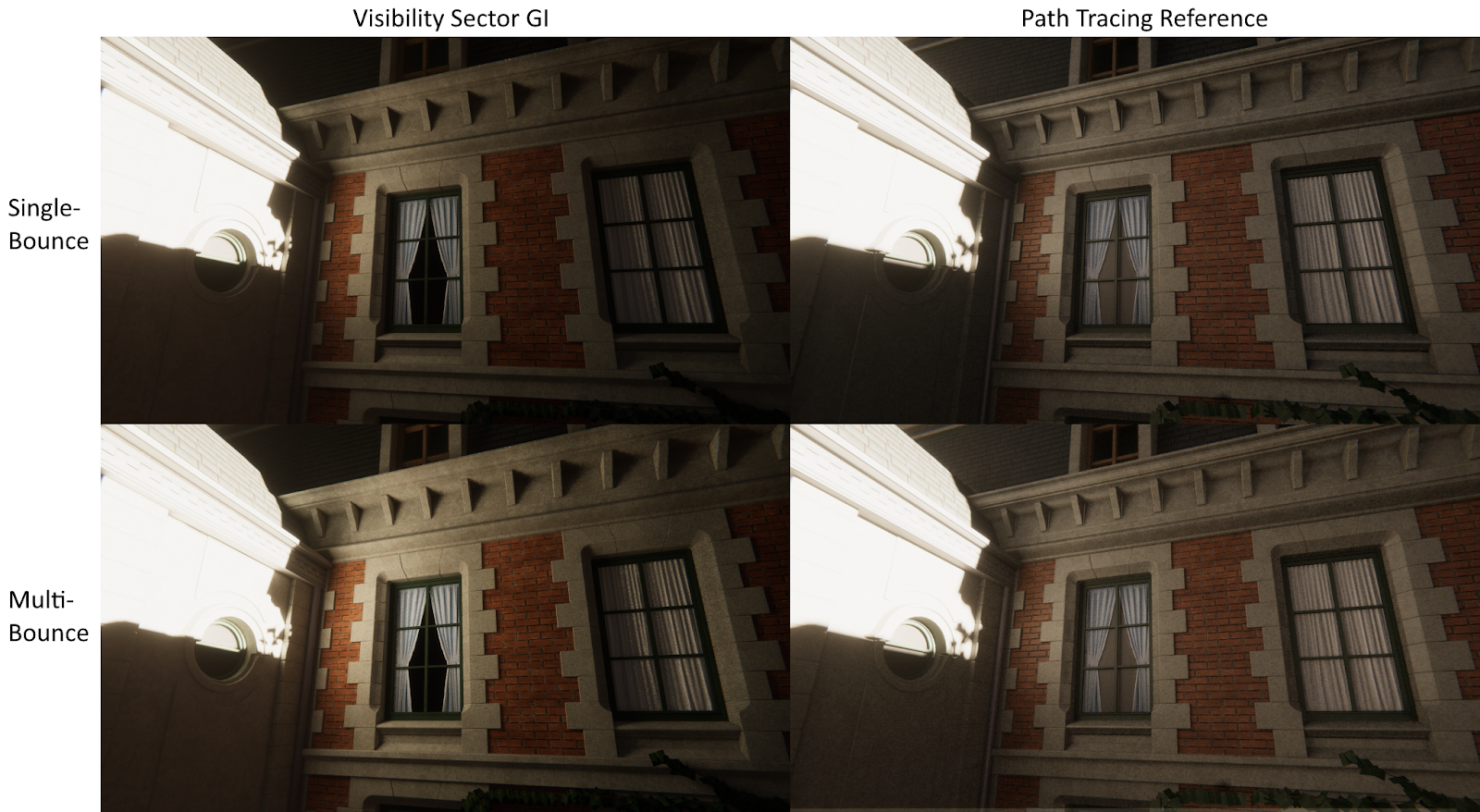}
  \caption{\label{fig:GIBounce}%
          Comparison of indirect diffuse algorithm and path tracing reference with single and multi-bounce indirect diffuse.}
\end{figure*}


\subsection{Indirect Diffuse}

In this subsection, we look at the visual quality and performance of the indirect diffuse portion of the algorithm. In addition to sampling the depth buffer, we also sample the HDR light buffer and the screen space normal buffer for every sample taken. The light buffer contains only the direct lighting (with shadows), as the ambient light is computed by our method. 
Figure \ref{fig:GIBounce} compares our result with a path tracing reference for single and multi-bounce indirect diffuse lighting. The direct lighting coming from the sun on the left wall bounces on the brick wall, illuminating it and casting indirect shadows. With multiple bounces, the light is even able to bounce back on the left wall, illuminating a shadowed region of the wall. Multiple bounces are implemented by injecting the indirect illumination into the light buffer to be used as input for the next frame. Light intensity needs to be properly balanced when using multi-bounce, or it can cause a feedback loop resulting in lighting accumulation over time. The result cannot match perfectly the path-traced reference since the algorithm operates only on the screen pixels as opposed to the entire scene geometry. One major drawback of our technique, along with screen-space methods, is that if direct light leaves the screen, the indirect lighting disappears.

\begin{figure}[htb]
  \centering
  \includegraphics[width=1.0\linewidth]{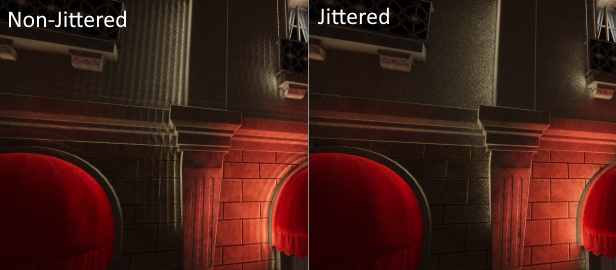}
  \caption{\label{fig:Jitter}
           When samples are evenly spaced along the sampling direction, a banding pattern can appear. Jittering the samples along the sampling direction helps mask the banding artifact.}
\end{figure}

\begin{figure*}[htb]
  \centering
  \includegraphics[width=.65\linewidth]{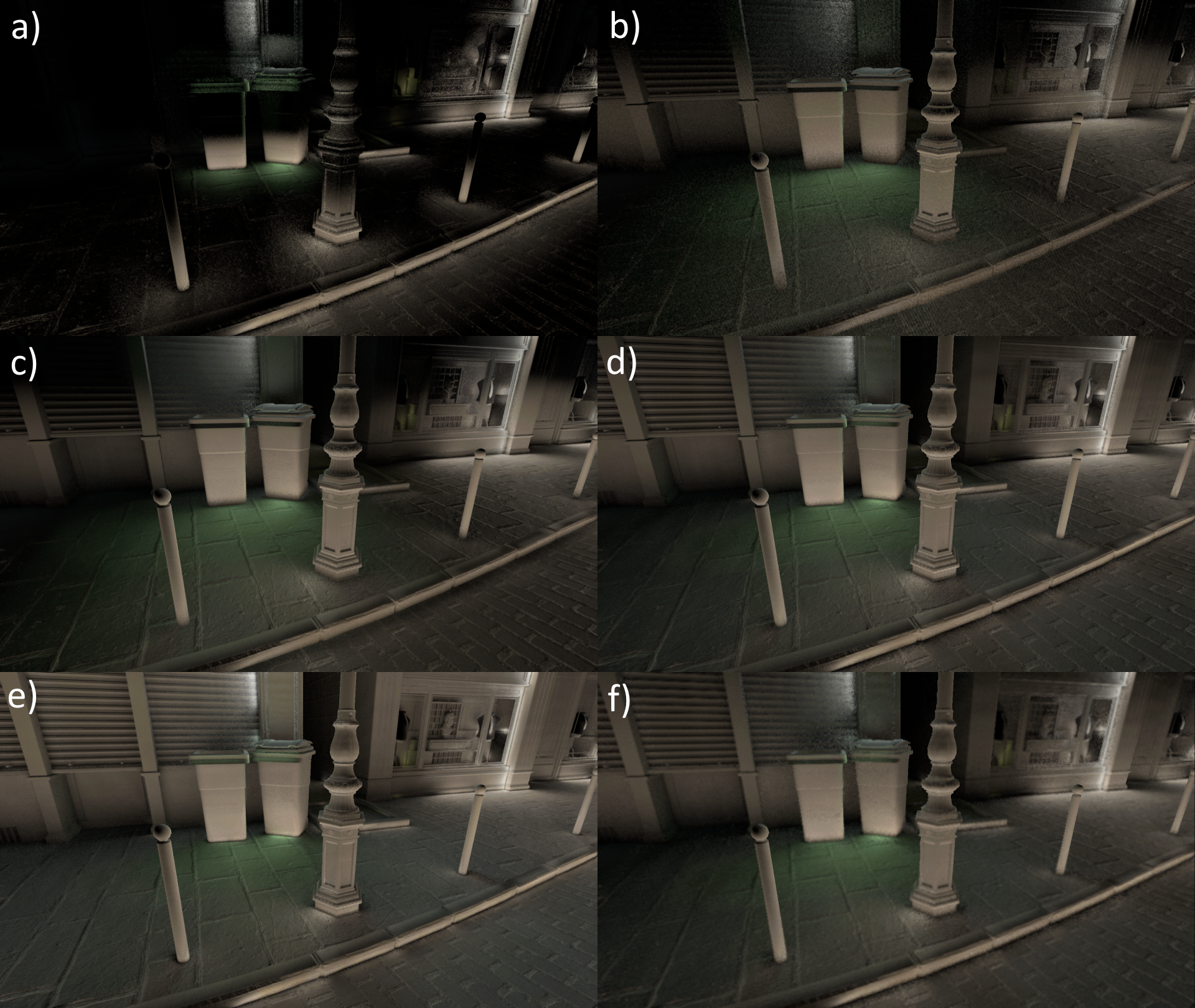}
  \caption{\label{fig:GIParam}%
          Resulting renders using different sample, radius, stepping, and resolution parameters for the indirect diffuse lighting. The parameters of each figure are indicated in Table \ref{table:2}.}
\end{figure*}

\begin{figure}[htb]
  \centering
  \includegraphics[width=1.0\linewidth]{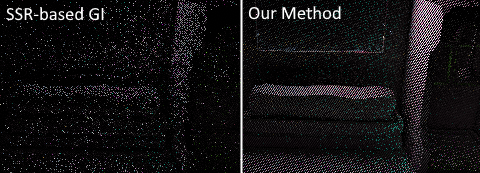}
  \caption{\label{fig:Noise}
           Left: 2 rays per pixel SSR tracing, with 16 steps per ray. There is a lot of noise, as a lot of rays exit the screen or pass behind surfaces without hitting anything. Right: Horizon search on both sides of the pixel using a visibility bitmask with 16 steps per horizon is a lot less noisy.}
\end{figure}


The number of samples taken has a big impact on quality. Having a too low sampling density can introduce banding artifacts. To mitigate the issue, samples are jittered along the sampling direction, as shown in Figure \ref{fig:Jitter}. 

Low sample density can also cause a loss of detail around small objects. To improve this, we distribute the samples exponentially around the shaded pixel, as nearby surfaces usually have more influence on the result than farther ones.

Figure \ref{fig:Noise} compares the amount of noise incurred when sampling the scene with SSR-like tracing and our method based on visibility bitmasks. Both techniques are taking the same maximum number of samples but the visibility bitmask approach has a lot less noise. Rays in SSR are doing at most one accumulation operation (when a hit is found). In contrast, the visibility bitmask approach is accumulating each sample that is visible from the current pixel.

This algorithm is bandwidth-intensive because we need to sample the HDR light buffer and the screen space normal buffer for every sample taken. Those sample locations are not correlated and impair caching. Table \ref{table:2} compares the performances of the indirect diffuse part of the algorithm with different sampling parameters. The corresponding renders are shown in Figure \ref{fig:GIParam}. Full resolution means that the shader is executed for every pixel of the final render resolution, whereas half-resolution executes it on a screen that is half the size (a quarter the number of pixels). The image is then upscaled with a classic bilateral upsampler to avoid aliasing. Rendering in half resolution is much more efficient (around 4x), but can introduce more flickering in the image and a blurrier result.

\begin{table}[h!]
\centering
\begin{tabular} { | m{10em} | m{1.2cm}| m{1.2cm} | m{1cm} | }
 \hline
 Configuration & Sampling & Denoising & Total \\ [0.5ex] 
 \hline\hline
 a) 8 samples, radius 1, const. steps, full res. & 0.9 ms & 0.33 ms & 1.23 ms \\ 
 \hline
 b) 8 samples, radius 4, const. steps, full res. & 1.7 ms & 0.33 ms & 2.03 ms \\
 \hline
 c) 16 samples, radius 4, const. steps, full res. & 2.3 ms & 0.33 ms & 2.63 ms \\
 \hline
 d) 16 samples, radius 4, exp. steps, full res. & 2.6 ms & 0.33 ms & 2.93 ms \\
 \hline
 e) 32 samples, radius 4, exp. steps, full res. & 4.0 ms & 0.33 ms & 4.33 ms \\
 \hline
 f) 16 samples, radius 4, exp. steps, half res. & 0.97 ms & 0.1 ms & 1.07 ms \\
 \hline
\end{tabular}
\caption{Render time at 1920x1080, with a constant thickness value of 0.2.}
\label{table:2}
\end{table}


\section{Conclusion and future work}

Previous screen space GI methods that rely on HBAO cannot handle thickness because of the assumption that the depth buffer is strictly a height field. Other techniques based on SSR tracing squander a lot of rays that end up passing behind and over surfaces, exiting the screen without hitting anything, thereby introducing a lot of noise. In contrast, our method combined the sampling efficiency of horizon-based methods, while retaining the capability of handling thickness properly along the way. The proposed algorithm is easy to understand and implement on modern GPUs and can be integrated into any horizon-based technique with only a mild performance overhead. Moreover, this method can also be used to improve ambient light sampling by taking into account the directionality of occlusion when integrating ambient light.

Even though the ray-tracing part of the algorithm handles thickness properly, since we operate on a single layer of the depth buffer, the actual object thickness is unknown. We are forced to rely on a constant thickness value that optionally increases linearly over the distance. A thickness heuristic that would give a plausible thickness per pixel would help improve the occlusion leaks around some very thin objects or light leaks behind very thick ones and would be interesting to explore in future work. At the performance level, SSGI can be an expensive algorithm because it is very demanding on GPU bandwidth and utilizes the cache poorly. Using a caching method similar to LSAO could potentially improve this. Finally, the common ambient light sources are either static (constant color, light probes) or expensive to update at runtime. We might investigate in the future ways to approximate low-frequency ambient irradiance dynamically.

\section{Acknowledgments and data statements}
Special thanks to Deepti Joshi (CDRIN) and Peter Shirley (NVIDIA) for their guidance in the redaction of this paper. We also want to thank our other colleagues namely Antoine Fortin (CDRIN), Olivier Leclerc (CDRIN), Steven Pigeon (UQAR), and Vahe Vardanyan (CDRIN) who have actively supported the current body of work. Most of the models are from the Amazon Lumberyard Bistro scene. This research is financed in part by the province of Quebec (Canada) via the grant "Programme d'aide à la recherche et au transfert (PART)". Data sharing not applicable to this article as no datasets were generated or analysed during the current study.

\bibliographystyle{eg-alpha-doi} 
\bibliography{Main}



\end{document}